\documentclass[twocolumn,showpacs,superscriptaddress, preprintnumbers , amsmath,amssymb,letterpaper]{revtex4-1}

\usepackage{lineno}
\usepackage{dcolumn}

\usepackage{bm}
\usepackage{amsmath}
\usepackage{multirow}
\usepackage{graphicx}
\usepackage{float}
\usepackage[colorlinks=true]{hyperref}

\begin{document}

\title{Equivalence of Gouy and Courant-Snyder phase}

\author{K. Floettmann}

\email{Klaus.Floettmann@DESY.de}

\affiliation{Deutsches Elektronen-Synchrotron, Notkestra\ss e 85, 22607 Hamburg, Germany}

\date{\today}

\begin{abstract}
The generation of electron vortex beams and the conversion of these beams into beams without angular momentum by means of astigmatic optical systems, or vice versa, has been pursued in the optical and the electron microscopy community, but also in the accelerator community in the past decades. Despite different conceptual approaches similar results have been achieved. By adapting the Courant-Snyder theory, which was originally developed for the description of optical properties of accelerators, to the description of laser modes it is shown, that identical mode converters have been developed for charged particle and for light beams, and that the Courant-Snyder phase and the Gouy phase are equivalent.  
\end{abstract}


\maketitle

\section{Introduction}
Electron vortex beams, i.e., electron beams carrying an angular momentum, have gained interest in the quantum optical and the electron microscopy communities in the past decade for their intriguing theoretical properties but also for their potential to enable new applications, ranging from the manipulation of molecules, clusters and nanoparticles or the probing of chiral structures up to the exploration of fundamental interactions in high energy collisions~\cite{Schattschneider2017}. Independently, and largely unnoticed by the quantum optical and electron microscopy communities, vortex beams have also been studied theoretically and experimentally in the accelerator community, where other techniques and applications have been developed. The apparent lack of communication and interaction between these communities finds a reason in the difference of basic concepts and notations, which obscures the equivalence of the underlying physical effects. Yet another reason is related to the fact that the very fundamental physical approach in these communities is very different and apparently exclusive. While one community is bound to wave mechanics and low quantum numbers the other is strictly related to point-like particle mechanics in the classical limit of very large quantum numbers. Still, similar results have been obtained in both physical realms and a detailed comparison leads to new insights relevant to both communities.\\
One of the techniques discussed in connection with vortex beams is a mode converter which transforms an angular momentum carrying Laguerre-Gaussian beam into an Hermite-Gaussian beam, which doesn't carry angular momentum, or vice versa. Originally proposed in the context of laser beams~\cite{Tamm1990, Abramochkin1991, Allen1992, Allen1993}, mode conversion of a charged beam in an electron microscopic is discussed by Schattschneider et al. in reference~\cite{Schattschneider2012} by adopting the wave mechanical approach developed for photon beams. This approach is further worked out in the recent publication~\cite{Kramberger2019}. Over a decade earlier a similar mode converter or beam adapter transforming a beam with angular momentum into a beam without angular momentum has already been introduced by Derbenev~\cite{Derbenev1998} in the context of electron cooling of proton beams. Besides thorough theoretical studies~\cite{Burov2000, Burov2002, Kim2003}  this led also to the proposal of a so-called flat-beam electron source~\cite{Brinkmann2001} and first experimental demonstrations of mode conversion at relativistic electron energies~\cite{Edwards2000, Edwards2001, Sun2004, Piot2006}.\\
The mode converters proposed in the various publications share superficial similarities, such as that in all cases cylindrical lenses are involved and that an incoming symmetric beam is transformed into an asymmetric beam, which is oriented at an angle of 45$^\circ$ with respect to the fundamental axis of the astigmatic lenses. Photon beam converters are generally made out of two lenses~\cite{Allen1993}, while electron beam converters require three quadrupole lenses~\cite{Kramberger2019, Burov2000, Burov2002, Kim2003, Brinkmann2001} for a general mode conversion. However, for the two lens scheme a specific distance is required between the lenses, while adjustable fields are conveniently applied in the three quadrupole lens adapter. So in each case three independent parameters need to be adjusted~\cite{footnote1}.\\   
Besides these obvious similarities two striking differences appear however. While in the context of lasers~\cite{Tamm1990, Abramochkin1991, Allen1992} and microscopy~\cite{Schattschneider2012, Kramberger2019} the discussion concentrates on the conversion of pure modes, a detailed mode description is not applied in the context of accelerators~\cite{Burov2000, Burov2002, Kim2003, Brinkmann2001}. The second, related difference is that in connection with the mode description the difference of the Gouy phase\cite{Gouy1890} in orthogonal planes of the astigmatic optical system is considered, while Derbenev's transformation is based on the difference of the Courant-Snyder phase advance~\cite{Courant1958} in the orthogonal planes of the mode converter. A central point in the discussion of the Derbenev transformation, yet missing in the discussion of lasers and microscopes, is however the preservation of the beam quality through the mode converter.\\
Below it will be shown that the Gouy phase and the Courant-Snyder phase advance are identical despite the fact that they are derived from very different conceptual approaches. This discussion requires a set of basic relations from both the wave mechanical photon and from the charged particle accelerator physics world, which will be introduced in the following sections. The presentation is extended at this point to highlight some general relations of the Courant-Snyder theory and the description of optical modes. In the last paragraph it is shown, that the conversion of pure modes leads to the same relations for the preservation of the beam quality as discussed in the field of accelerators, which demonstrates that the Derbenev transformation and the pure mode converter are identical.

\section{Beam quality and envelope equation in a drift}
A Cartesian coordinate system is assumed in which the $z$ direction corresponds to the longitudinal coordinate of the beam. Unless explicitly mentioned, all equations which are derived for the transverse $x$ direction are equally valid for the transverse y direction.\\
A particle beam consists of an ensemble of point like particles traveling predominantly in the $z$ direction. For a constant number of particles, i.e., no particle loss, the beam emittance is a measure of the beam quality and a conserved quantity of motion in a linear, uncoupled transport system without chromatic effects, i.e., free space optics consisting of free propagation sections (drifts) and linear focusing elements (lenses). The transverse emittance is defined as
\begin{equation}
{\varepsilon _x} = \sqrt {\left\langle {{x^2}} \right\rangle \left\langle {{{x'}^2}} \right\rangle  - {{\left\langle {xx'} \right\rangle }^2}},
\label{1.1}
\end{equation}
where the prime indicates the derivative w.r.t. the longitudinal coordinate $x' = \frac{d}{{dz}}x$ and $\left\langle {} \right\rangle $ defines the second central moment of a particle distribution. The smaller the emittance the better can a beam be focused and the smaller stays the beam divergence. The index $x$ at ${\varepsilon _x}$ will be omitted in the following. Assuming for simplicity a normalized distribution in the transverse phase-space $\rho  = \rho \left( {x,x'} \right)$, $\int {\rho \;dxdx' = 1} $ the second central moments read in integral form as
\begin{equation}
\begin{gathered}
  \left\langle {{x^2}} \right\rangle  = \int {\int {\rho {x^2}dxdx'} }  - {\left( {\int {\int {\rho x\;dxdx'} } } \right)^2} \hfill \\ 
  \left\langle {{{x'}^2}} \right\rangle  = \int {\int {\rho {{x'}^2}dx'dx} }  - {\left( {\int {\int {\rho x'dx'dx} } } \right)^2} \hfill \\ 
  \left\langle {xx'} \right\rangle  = \int {\int {\rho xx'dx'dx} }  - \int {\int {\rho x\;dxdx'} } \int {\int {\rho x'dx'} } dx. \hfill 
\end{gathered}
\label{1.2}
\end{equation}
All integrals range from minus to plus infinity.\\
The square root of the second central moment defines rms quantities, which are in general labeled with a $\sigma $.\\
\begin{figure}[htb]
\centering
    \includegraphics*[width=80mm]{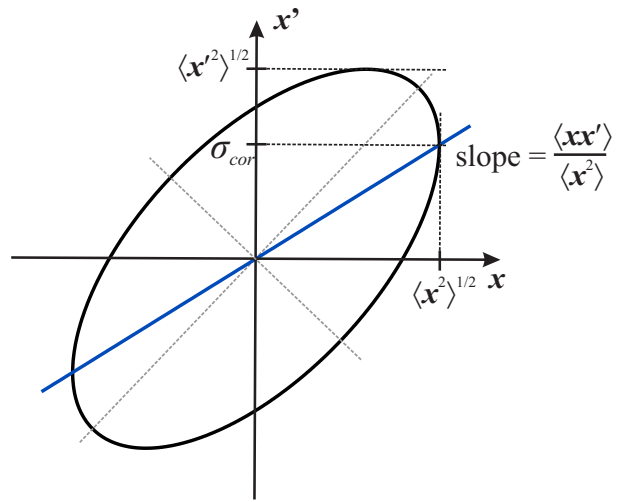}
    \caption{Transverse phase-space and rms ellipse with correlation straight (blue) and some relevant quantities.}
\label{fig.1}
\end{figure}
Figure~\ref{fig.1} sketches the transverse phase-space with an rms ellipse. It can be shown, that the rms quantities of arbitrary particle distributions form an ellipse in phase-space independently of all details of the distribution~\cite{Floettmann2003}. The emittance is proportional to the area $A$ of the ellipse $A={\pi\varepsilon}$. The orientation of the ellipse is described by the correlation straight, which follows a least square fit to the phase-space distribution. Thus it is not a principal axis of the ellipse, but a symmetry axis in the sense that the distance to the upper and the lower branch of the ellipse is equal. The slope of the straight is given by $\left\langle xx' \right\rangle/\left\langle x^2 \right\rangle$. A negative slope of the correlation straight describes a converging beam, a positive slope, as in the figure, corresponds to a divergent beam. At a focus position the principal axes of the ellipse are aligned to the coordinate system and the correlation term $\left\langle xx'\right\rangle$  is zero. Eq.~\ref{1.1} reduces here to
\begin{equation}
\varepsilon  = \sqrt {\left\langle {x_0^2} \right\rangle \left\langle {x'_0}^2 \right\rangle }  = {\sigma_0}{\sigma'_0},
\label{1.3}
\end{equation}
where the index 0 is introduced to indicate the focus position. ${\sigma '_0}$ is called far-field diffraction angle in light optics.\\
The emittance can be related to the Liouville phase-space volume, but there are important differences.\\
The Liouville theorem states that the 6-dimensional canonical phase-space density along trajectories of the system is time independent (for a Hamiltonian system and non-interacting particles). For a given ensemble of particles one can define a phase-space volume, occupied by the ensemble, by integrating over all phase-space coordinates. This volume is a constant of motion. The form of this volume is however unspecified and can change, while the emittance is related to an ellipse or, in higher dimensions, to a hyper-ellipsoid. When the particle ensemble passes for example through a non-linear focusing field, the Liouville phase-space will change its form but not its volume. The emittance in contrast will increase. The emittance is thus more sensitive to detrimental effects than the Liouville phase-space volume and therefore the relevant measure of the beam quality. A non-linear focusing field is also an example which leads to a correlated (higher-order) distortion of the phase-space. Since it is not fundamentally excluded to compensate correlated distortions, i.e., by another non-linear field in this example, the emittance can also become smaller, if a correlated distortion exists in the phase-space.\\ 
In general we can split the emittance into a statistical and a correlated term as
\begin{equation}
{\varepsilon ^2} = \varepsilon _{st}^2 + \varepsilon _{cor}^2.
\label{1.4}
\end{equation}
The statistical emittance is closely related to the Liouville phase-space volume and represents the minimal possible emittance of the phase-space under consideration, while the correlated emittance contribution -- if unequal to zero -- might be compensated in an appropriate optical configuration.\\
Correlated phase-space distortions appear in manifold forms, besides non-linearities, correlations between (ideally uncorrelated) coordinates are to be mentioned. The angular momentum of a beam is such a contribution since it is a correlation between angle and space coordinates of orthogonal degrees of freedom, i.e. $\left\langle xy'\right\rangle \ne 0$ and $\left\langle yx'\right\rangle\ne 0$. As will be shown below this correlated contribution of the beam emittance is compensated in a mode converter.\\
Eq.~\ref{1.3} is up to a factor of 4 equal to the beam parameter product ($BPP$) usually defined in laser physics. The so-called M-square value ${\textbf{M}^2}$ is related to the beam parameter product and thus to the beam emittance by
\begin{equation}
BPP = {\textbf{M}^2}\frac{\lambda }{\pi } = \frac{{2{\textbf{M}^2}}}{k} = 4{\sigma _0}{\sigma '_0} = 4\varepsilon
\label{1.5}
\end{equation}
${\lambda}$ is the wavelength of the laser beam and $k$ its wave number.\\
Note, that also for a laser beam the transverse beam size is defined by the rms size or two times the rms size and not by the FWHM or the $1/e$ width. The advantage of the rms size in comparison to other beam size definitions is that it leads to simple equations describing the beam envelope independently of the details of the particle distribution~\cite{Floettmann2003} or wave field under consideration~\cite{Siegman1991}.\\
As simplification, the square in the notation of the M-square value is dropped in the following discussion, i.e. $M={\textbf{M}^2}$, so that especially $M^2=({\textbf{M}^2})^2$.\\
The variation of the rms beam size of a particle distribution traveling through a piece of beam line follows from the spatial derivatives of the second moment with respect to $z$ as~\cite{Sacherer1971}:
\begin{equation}
\sigma  =
\left\langle x^2 \right\rangle^\frac{1}{2}
\label{1.6}
\end{equation}
\begin{equation}
\sigma _{cor} = {(\sigma)^\prime} = \frac{d}{d z}{\left\langle x^2 \right\rangle}^\frac{1}{2} = \frac{\left\langle {xx'} \right\rangle } {{\left\langle x^2 \right\rangle}^\frac{1}{2}}
\label{1.7}
\end{equation}
\begin{equation}
(\sigma)^{\prime \prime} = \frac{d ^2}{d {z^2}}{\left\langle x^2 \right\rangle ^\frac{1}{2}} = \frac{\left\langle {x'^2} \right\rangle }{{\left\langle {{x^2}} \right\rangle }^\frac{1}{2}} + \frac{\left\langle xx'' \right\rangle }{{\left\langle {{x^2}} \right\rangle}^\frac{1}{2}} - \frac{{\left\langle {xx'} \right\rangle}^2}{{\left\langle x^2 \right\rangle}^\frac{3}{2}}.
\label{1.8}
\end{equation}
Here it is implicitly assumed, that the phase-space density is constant $\frac{d}{dz}\rho  = 0$, 
 i.e., that the Liouville theorem holds.\\
Note the difference of correlated divergence ${\sigma _{cor}}$ and rms divergence $\sigma ' = {\left\langle x'^2 \right\rangle ^\frac{1}{2}}$, cf. Figure~\ref{fig.1}.\\
From the definition of the emittance (Eq.~\ref{1.1}) follows
\begin{equation}
\left\langle x'^2 \right\rangle=\frac{\varepsilon ^2}{\left\langle x^2 \right\rangle } + \frac{{\left\langle {xx'} \right\rangle }^2}{\left\langle x^2 \right\rangle},
	\label{1.9}
	\end{equation}
which is used to replace $\left\langle {{x'^2}} \right\rangle $ in Eq.~\ref{1.8} which leads to the differential form of the rms envelope equation:
\begin{equation}
{\left( \sigma  \right)^{\prime \prime }} = \frac{\varepsilon ^2}{{\left\langle x^2 \right\rangle }^\frac{3}{2}} + \frac{\left\langle xx'' \right\rangle}{{\left\langle x^2 \right\rangle }^\frac{1}{2}}.
\label{1.10}
\end{equation}
Note that no assumptions about the details of the particle distribution have been made in deriving Eq.~\ref{1.10}. It describes hence the rms beam size for arbitrary distributions as long as the emittance stays constant.\\
In a drift $x''=0$ and a general solution of the envelope equation reads as:
\begin{equation}
{\sigma ^2}(z) = \sigma _{z = 0}^2 + 2{\sigma _{z = 0}}{\left( {\sigma _{z = 0}} \right)^\prime }z + \left( {\frac{{\varepsilon ^2}}{\sigma _{z = 0}^2}+ {{\left( {{\sigma _{z = 0}}} \right)}^\prime }^2} \right){z^2}.
\label{1.11}
\end{equation}
Note that ${\sigma _{z = 0}}{\left( {{\sigma _{z = 0}}} \right)^\prime } = {\left\langle {xx'} \right\rangle _{z = 0}}$ represents the linear correlation term in phase-space, cf. Figure~\ref{fig.1}.\\
By setting $z=0$ at a focus position and thus to a position where $ \left( {\sigma _{z = 0}} \right)^\prime=0 $ Eq.~\ref{1.11} simplifies to
\begin{equation}
{\sigma ^2}(z) = \sigma _0^2 + \frac{{{\varepsilon ^2}}}{{\sigma _0^2}}{z^2}
\label{1.12}
\end{equation}
Equivalent equations to Eq.~\ref{1.11} and Eq.~\ref{1.12} are derived in~\cite{Siegman1991} for a general photon beam. Especially the equivalent formula to Eq.~\ref{1.12} as used in light optics is the well-known relation
\begin{equation}
{\sigma ^2}(z) = \sigma _0^2\left( {1 + \frac{{{z^2}}}{{Z_R^2}}} \right),
	\label{1.13}
	\end{equation}
with the Rayleigh length ${Z_R}$ defined as
	\begin{equation}
	{Z_R} = \frac{{4\pi \sigma _0^2}}{{M\lambda }} = \frac{{\sigma _0^2}}{\varepsilon }.
	\label{1.14}
	\end{equation}

\section{Optical systems and Courant-Snyder formalism}

\begin{figure*}[ttt]
\centering
\includegraphics*[width=\textwidth]{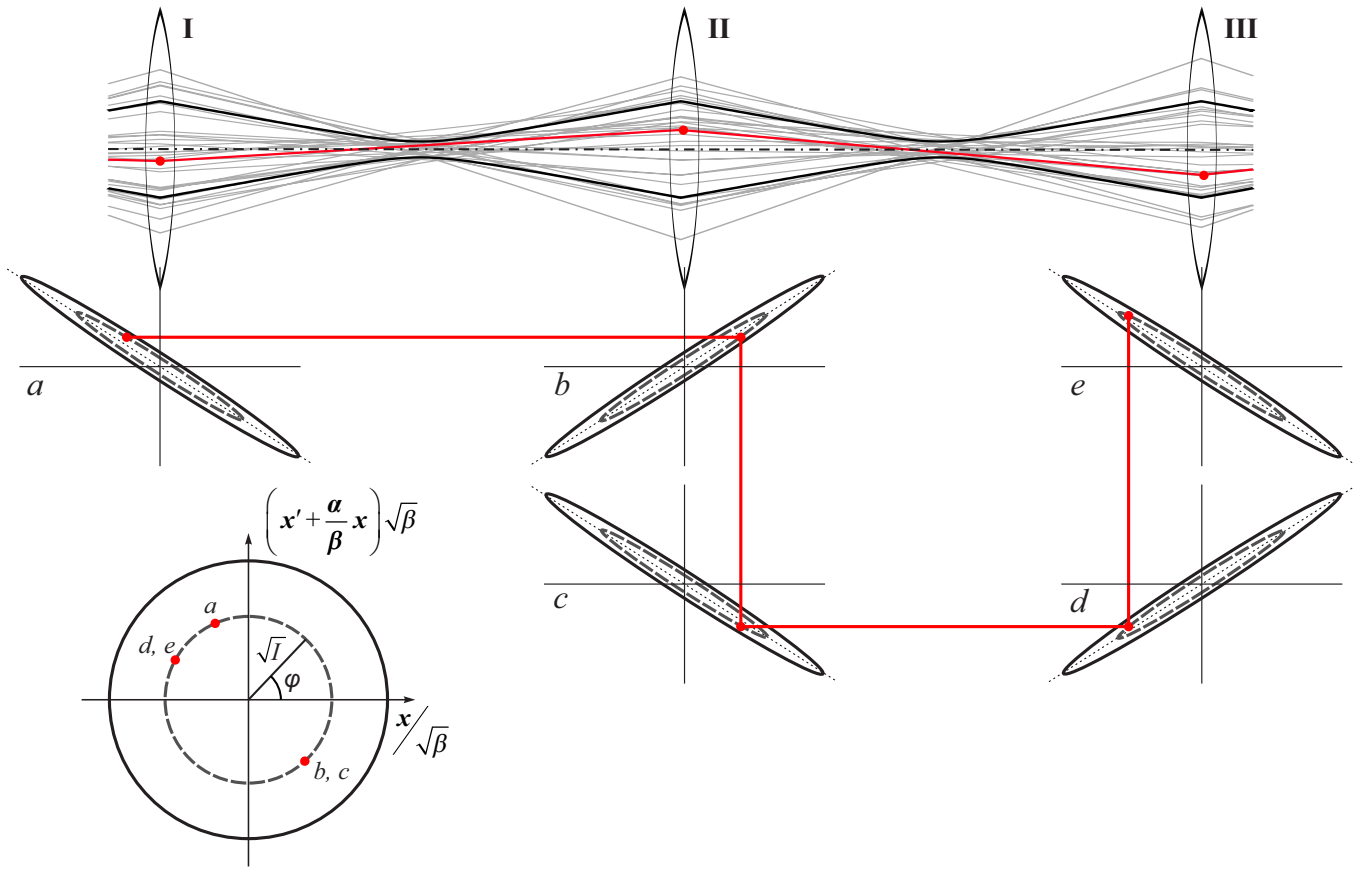}
    \caption{Graphical illustration of the phase advance in a periodical optical system. At the top a number of particle trajectories or optical rays are shown in gray together with the rms beam envelope (black) and three lenses. A single trajectory is highlighted in red. Below this beam line the corresponding development of the phase-space is plotted, starting right at the exit of lens \textbf{I} with plot \textit{a}. The basic motions in phase-space are shearing operations. In a drift the trajectory angle stays constant and the position of each particle develops as ${x_{end}} = {x_{initial}} + x'z$. The convergent beam of plot a develops thus into a focused beam (not shown) and further into a divergent beam in \textit{b}. In the ideal thin lens, the position is constant and the angle develops as ${x'_{end}} = {x'_{initial}} - Kx$, which transforms the phase-space into the state shown in \textit{c}. After passing another drift and the third lens the phase-space is in the same orientation as in the beginning, just rotated by 360$^\circ$. The red line follows the highlighted particle in phase-space. It is seen that the particle stays on its own ellipse which has the same orientation and eccentricity as the rms ellipse. This is the case for each particle and thus implies that the number of particles or the total energy encircled by such an ellipse is constant. Note that the particle does however not reach its starting position in plot \textit{e}. The difference is described by the phase advance which is 322$^\circ$ in this example. For the sub-panel in the lower left the Courant-Snyder parameters have been used to transform the rms ellipse to a circle with a radius corresponding to the square root of the emittance. Here the radius and the rotation angle of individual particles correspond to the action-phase variables of the Hamiltonian.}
\label{fig.2}
\end{figure*}
So far only a simple drift has been considered. A general linear optical transport system consists of focusing and defocusing lenses and drifts in between. A lens is described to first order by $x'' =  - Kx$ with the positive focusing or negative defocusing strength K or inverse focal length $1/f=K$. For an optical system, $K(z)$ is a discontinuous function combing all elements, i.e., $K(z) \ne 0$ in lenses and $K(z) = 0$ in all drifts. Substituting $x'' =  - Kx$ in Eq.~\ref{1.10} leads to the differential envelope equation for this case as:
\begin{equation}
{\left( \sigma  \right)^{\prime \prime }} + K(z)\sigma  = \frac{{{\varepsilon ^2}}}{{{\sigma ^3}}}.
	\label{1.15}
	\end{equation}
Eq.~\ref{1.15} describes the development of the beam size, i.e., the second central moment through a system of lenses and drifts.\\ 
An equivalent relation for the average beam position or the first moment reads:
\begin{equation}
	\bar x'' + K(z)\bar x = 0.
	\label{1.16}
	\end{equation}
Here the bar is used for the direct average, i.e., $ \bar x=\int {\int {\rho xdxdx'}}$, while the angle brackets define the central average (Eq.~\ref{1.2}).
The trajectory of a single particle through the beam line follows the same equation as the average position of an ensemble of particles; Eq.~\ref{1.16} is thus valid for both cases.\\
The Courant-Snyder theory describes a general solution of Eqs~\ref{1.15} and \ref{1.16} and the relation between them. In the following, only those aspects of the theory which are required to introduce the phase advance are discussed. Only a simple system without dispersion, coupling etc. is considered. For a detailed discussion of the Courant-Snyder theory the reader is pointed to reference~\cite{Courant1958} or a textbook on accelerator physics.\\
Originally developed in context of ring type accelerators, where the focusing function $K(z)$  is periodic, the Courant-Snyder formalism can also be applied to finite transport lines, which can be viewed as being segments of larger periodic systems. The main difference between periodic systems and beam lines is that the periodic case leads to a unique solution, while the solution for a transport line depends on the initial conditions.\\ 
Rather than describing the beam properties as the transverse beam size directly, the Courant-Snyder theory leads to solutions in the form of optical functions. Most prominent is the ${\beta}$-function which is related to the beam size by
\begin{equation}
\beta (z) = \frac{{{\sigma ^2}(z)}}{\varepsilon }.
\label{1.17}
\end{equation}
The other optical functions are defined by $\alpha (z) =  -\beta '/2$ and $\gamma (z) = {\sigma '}^2/ \varepsilon$. While $\beta $ and $\gamma $ are related to rms beam size and rms divergence, $\alpha $ is related to the correlation term by $\alpha  =  - \frac{{\left\langle {xx'} \right\rangle }}{\varepsilon }$. The relation $\beta \gamma  - {\alpha ^2} = 1$ holds and the quantity
\begin{equation}
I = \gamma {x^2} + 2\alpha xx' + \beta {x'^2}
\label{1.18}
\end{equation}
is a constant of motion.\\

Eq.~\ref{1.18} is a coordinate representation of an ellipse. For particle coordinates on the rms ellipse $I=\varepsilon$ holds. The optical functions are normalized to the beam emittance and are hence valid for beams with arbitrary emittance. Thus they describe the characteristics of an optical system independent of the beam quality and allow to separate the influence of the optics and of the beam quality on the local beam properties.\\
At a focus, the ${\beta}$-function, ${\beta_0}$, corresponds to the Rayleigh length, Eq.~\ref{1.14}, and near a beam waist it develops as (cf. Eq.~\ref{1.12})
\begin{equation}
\beta (z) = {\beta _0} + \frac{{{z^2}}}{{{\beta _0}}},
\label{1.19}
\end{equation}
where again ${z_0} = 0$ is assumed.\\
Solving Eq.~\ref{1.17} for $\sigma $, calculating the derivatives w.r.t. $z$ and introducing all into Eq.~\ref{1.15}, transforms the differential equation for the beam size into a differential equation for the ${\beta}$-function
\begin{equation}
	\frac{{\beta \beta ''}}{2} - \frac{{{{\beta '}^2}}}{4} + K{\beta ^2} = 1.
	\label{1.20}
	\end{equation}
Also the relation for the first moment, Eq.~\ref{1.16}, needs to be transformed into the system of new variables. Noting that Eq.~\ref{1.16} has the form of Hills differential equation (a linear differential equation without first derivative term), motivates to seek for a solution of the form
	\begin{equation}
	\bar x = \sqrt {I\beta } {e^{i\phi }}.
	\label{1.21}
	\end{equation}
While the amplitude $\sqrt I $ (Eq.~\ref{1.18}) is a constant of motion the phase $\phi $ is $z$-dependent just as $\beta $.\\
Eqs~\ref{1.21} and \ref{1.20} form a new system of equations describing the first and second moment of a distribution in a linear transport system. Still missing is a relation to determine the phase $\phi $. Introducing the derivatives of Eq.~\ref{1.21} into Eq.~\ref{1.16} leads to the relation:
	\begin{equation}
	\Re \left\{ {\frac{{\bar x''+K \bar x}}{{\sqrt I{e^{i\phi }}}}} \right\} = \frac{{\beta \beta ''}}{2} - \frac{{{{\beta '}^2}}}{4} + K{\beta ^2} - {\beta ^2}{\phi '^2} = 0,
	\label{1.22}
	\end{equation}
where only the real part is required at this point.\\ 
From Eq.~\ref{1.20} and Eq.~\ref{1.22} follows
	\begin{equation}
	\phi  = \int {\frac{1}{\beta }} \;dz
	\label{1.23}
	\end{equation}
	Eq.~\ref{1.23}, which fulfills also the imaginary part of Eq.~\ref{1.22}, defines the Courant-Snyder phase advance. The equation describes the phase difference between two points of a beam line. An appropriate initial phase can be added, however, in practice often only the phase advance between two points is relevant.\\ 
Figure~\ref{fig.2} illustrates the phase advance and relations to other quantities. The Courant-Snyder theory is a versatile and powerful formalism to describe and design optical systems. The phase advance is for instance the key parameter to study and improve the stability of optical solutions as required to store a beam in a ring, or to propagate errors through optical systems or to solve imaging problems. Some illustrative examples are summarized in the Appendix.

\section{Mode description and Gouy phase}
Modes constitute a complete and orthogonal set of functions describing solutions of the scalar wave equation (in paraxial approximation). Arbitrary field distributions can be expanded in terms of these modes. The following discussion concentrates on Hermite-Gauss modes, the results can however also be transferred to other basis as for example to the Laguerre-Gauss basis.\\ 
The transverse electric field of a Hermite-Gauss mode reads in standard notation as
	\begin{equation}
	E \propto \frac{1}{w}{H_m}\left( {\sqrt 2 \frac{x}{w}} \right){H_n}\left( {\sqrt 2 \frac{y}{w}} \right){e^{\left( { - \frac{{{x^2} + {y^2}}}{{{w^2}}}} \right)}}{e^{i\theta }}.
	\label{1.24}
	\end{equation}
${H_m}$ and ${H_n}$ are Hermite polynomials of order $m$ and $n$ and $w = w(z)$ describes the development of the beam size as function of the longitudinal coordinate. Thus a Hermite-Gauss mode is essentially a product of Hermite polynomials -- one for each transverse degree of freedom -- multiplied by a Gaussian density distribution.\\
The phase $\theta $ will be discussed below. First we concentrate on the beam size, which is determined by the real part of Eq.~\ref{1.24}.\\ 
In general $w$ is just named beam size but without specifying how this beam size is defined. It is indeed the beam size of the fundamental Gaussian mode ($m = n = 0$), but not the beam size of the higher modes. Calculating the rms beam sizes from Eq.~\ref{1.24} yields
	\begin{equation}
	\begin{gathered}
  {\sigma _x} = \sqrt {2m + 1} \frac{w}{2} \hfill \\
  {\sigma _y} = \sqrt {2n + 1} \frac{w}{2}. 
\end{gathered} 
\label{1.25}
\end{equation}
Since the beam size scales everywhere with $\sqrt {2m + 1} $ and $\sqrt {2n + 1}$, respectively, also the far-field diffraction angle ${\sigma '_0}$ scales with this factor and thus the beam quality factors follow as
	\begin{equation}
	\begin{gathered}
  {M_x} = 2m + 1 \hfill \\
  {M_y} = 2n + 1. 
\end{gathered} 
\label{1.26}
\end{equation}
Eq.~\ref{1.25} suggests to interpret $w$ in terms of the ${\beta}$-function and set $w = 2\sqrt {\frac{\beta }{{2k}}} $, so that the field acquires the form (constants have been dropped)
	\begin{equation}
	E \propto \sqrt {\frac{1}{\beta }} {H_m}\left( {\sqrt {\frac{k}{\beta }} x} \right){H_n}\left( {\sqrt {\frac{k}{\beta }} y} \right){e^{\left( { - \frac{k}{\beta }\frac{{\left( {{x^2} + {y^2}} \right)}}{2}} \right)}}{e^{i\theta }}
	\label{1.27}
	\end{equation}
The rms beam sizes read now as
	\[\begin{gathered}
  {\sigma _x} = \sqrt {\frac{{2m + 1}}{{2k}}\beta }  = \sqrt {{\varepsilon _x}\beta }  \hfill \\
  {\sigma _y} = \sqrt {\frac{{2n + 1}}{{2k}}\beta }  = \sqrt {{\varepsilon _y}\beta } 
\end{gathered} \]	(1.28)
in accordance to Eq. (1.17).\\
The standard notation of modes describes the modes thus not with a mode number independent beam size, but with a mode number independent ${\beta}$-function. Or, in other words, the Rayleigh length (= ${\beta}$-function at the focus) and the Courant-Snyder phase advance is independent of the mode number. It has to be noted that the orthogonality of the modes requires that the arguments of the polynomials and of the exponential term in each transverse degree of freedom, are identical for all modes. The decomposition of a field distribution into orthogonal modes requires thus a description in accordance to Eq.~\ref{1.27}.\\
The phase term reads in explicit form
	\begin{equation}
	\theta  = kz + \frac{{\varsigma \left( {{x^2} + {y^2}} \right)}}{{w_0^2(1 + {\varsigma ^2})}} - \omega t - \left( {m + n + 1} \right){\varphi _G},
	\label{1.29}
	\end{equation}
with $\varsigma  = \frac{z}{{{Z_R}}} = \frac{z}{{{\beta _0}}}$ and ${z_0} = 0$.\\ 
${\varphi _G}$ is an additional phase advance which a wave going through a focus obeys in comparison to an unfocused wave. It was first observed by L. Gouy in 1890 with a light beam and later explained by him on the basis of Huygens principle. It follows thus from basic physics principles and should not be viewed as an anomaly. The Gouy phase is given by
	\begin{equation}
	{\varphi_G} = \arctan \varsigma, 
	\label{1.30}
	\end{equation}
where a round beam focus is assumed. For $\varsigma $ ranging from minus infinity to infinity the Gouy phase is equal to $\pi $.\\
To show that the Courant-Snyder phase Eq.~\ref{1.23} is a generalized form of the Gouy phase the arc tangent needs to be written in integral form as:
	\begin{equation}
	{\varphi _G} = \arctan \varsigma  = \int {\frac{1}{{1 + {\varsigma ^2}}}\;} d\varsigma, 
	\label{1.31}
	\end{equation}
with $d\varsigma  = \frac{{dz}}{{{\beta _0}}}$. Using Eq.~\ref{1.19} to replace $\varsigma  =\frac{z}{\beta _0}$ leads to
	\begin{equation}
	{\varphi _G} = \int {\frac{1}{\beta }\;} dz.
	\label{1.32}
	\end{equation}
The integral term is the Courant-Snyder phase advance, Eq.~\ref{1.23}.\\
For the general astigmatic case the Gouy phase needs to be separated into a horizontal and a vertical part. With
	\begin{equation}
	m + n + 1 = \frac{{{M_x}}}{2} + \frac{{{M_y}}}{2}
	\label{1.33}
	\end{equation}
we can write
	\begin{equation}
	\left( {m + n + 1} \right){\varphi _G} = \frac{{{M_x}}}{2}{\phi _x} + \frac{{{M_y}}}{2}{\phi _y},
	\label{1.34}
	\end{equation}
where ${\phi _x}$ and ${\phi _y}$ denote the Courant-Snyder phase advance in the horizontal and the vertical plane. Note, that the ${\beta}$-function and the focus position may differ for both planes in this representation. Eq.~\ref{1.34} displays the well-known result that the Gouy phase of an astigmatic focus (from minus infinity to infinity) is only $\frac{\pi}{2}$.\\
The second term of interest in Eq.~\ref{1.29}, $\frac{{\varsigma \left( {{x^2} + {y^2}} \right)}}{{w_0^2(1 + {\varsigma ^2})}}$, defines parabolas of constant phase in the transverse coordinates. The apex curvature radius $R$ of the constant phase lines or phase front is given by $R = z\left( {1 + {\varsigma ^{ - 2}}} \right)$. The phase curvature term can thus also be written in the form 
\begin{equation}
\frac{{\varsigma \left( {{x^2} + {y^2}} \right)}}{{w_0^2(1 + {\varsigma ^2})}} =   \frac{{k\left( {{x^2} + {y^2}} \right)}}{{2R}}.
\label{1.34a}
\end{equation}
Eq.~\ref{1.34a} describes the development of the phase curvature in a drift, here $\alpha  =  -\frac{\beta '}{2} =  -\frac{z}{\beta_0} =  - \varsigma $ holds. With this relation and the substitution as described above the term can also be written as
	\begin{equation}
	\frac{{\varsigma \left( {{x^2} + {y^2}} \right)}}{{w_0^2(1 + {\varsigma ^2})}} =   \frac{{k\left( {{x^2} + {y^2}} \right)}}{{2R}} =  - k\frac{\alpha }{\beta }\frac{{{x^2} + {y^2}}}{2}.
	\label{1.35}
	\end{equation}
A quadratic phase term corresponds to a linear correlation of transverse angles and positions. (A phase can be written as product of wave number and position, thus ${x^2} \propto {k_x}(x)x$ where ${k_x}$ depends linearly on x. Since $x' =\frac{k_x}{k_z}$ holds the divergence depends linearly on $x$.) The radius of curvature is thus related to the correlation straight of the phase-space, cf. Figure~\ref{fig.1} and \cite{Siegman1991}:
	\begin{equation}
	\frac{{\left\langle {xx'} \right\rangle }}{{\left\langle {{x^2}} \right\rangle }} =  - \frac{\alpha }{{\beta }} =   \frac{1}{{R}}
	\label{1.36}
	\end{equation}
Also this term can now be separated into the contributions of the two orthogonal planes and thus Eq.~\ref{1.24} can be written as product of two functions, each of which depends only on parameters of one transverse coordinate
	\begin{equation}
	\begin{gathered}
  E \propto {F_x}\left( {x,{\beta _x},{\alpha _x},m} \right){F_y}\left( {y,{\beta _y},{\alpha _y},n} \right) \hfill \\ 
  {F_x} = \frac{1}{{\beta _x^\frac{1}{4}}}{H_m}\left( {\sqrt {\frac{k}{{{\beta _x}}}} x} \right){e^{\left( { - \frac{{k{x^2}}}{{2{\beta _x}}}} \right)}}{e^{i\frac{1}{2}\left( {k\left[ {z - \frac{{{\alpha _x}}}{{{\beta _x}}}{x^2}} \right] - \omega t - {M_x}{\phi _x}} \right)}} \hfill \\ 
  {F_y} = \frac{1}{{\beta _y^\frac{1}{4}}}{H_n}\left( {\sqrt {\frac{k}{{{\beta _y}}}} y} \right){e^{\left( { - \frac{{k{y^2}}}{{2{\beta _y}}}} \right)}}{e^{i\frac{1}{2}\left( {k\left[ {z - \frac{{{\alpha _y}}}{{{\beta _y}}}{y^2}} \right] - \omega t - {M_y}{\phi _y}} \right)}}. 
\end{gathered} 
\label{1.37}
\end{equation}
Eq.~\ref{1.37} is the fully astigmatic description of Hermite-Gaussian modes. Just like Eq.~\ref{1.27}, which requires equal optical functions in orthogonal planes, it is an exact solution of the paraxial Helmholtz equation. 

\section{Mode conversion}
In the flat-beam electron source~\cite{Brinkmann2001}, the initial beam is generated by immersing the cathode in a solenoid field. As stated by the Busch theorem~\cite{Busch1926} this leads to a free vortex beam in which the electrons have intrinsic angular momenta ranging from zero up to a large absolute value related to the conditions at the cathode. Details including the quantum mechanical treatment of the Busch theorem are discussed in a parallel paper to this publication~\cite{Floettmann2020}. This Laguerre-Gaussian-like beam is converted into a Hermite-Gaussian-like beam which is characterized by a large asymmetry in the beam quality and the cancellation of the coupling terms in the transport matrices. Fundamental for the treatment is the beam quality in the two transverse planes before and after the converter. A detailed discussion in terms of the electron beam emittance is presented by Kim in ref.~\cite{Kim2003}. In the following it will be shown that the mode description leads to the same results.\\
The mode converter is conveniently discussed by comparing the angular spectrum of Hermite-Gauss modes with that of Laguerre modes. The angular spectrum of Hermite-Gauss modes is given by~\cite{Pampaloni2004}
	\begin{equation}
	\tilde E \propto \int\limits_{ - \infty }^\infty  {d{k_x}\int\limits_{ - \infty }^\infty  {dk_y} \;k_x^mk_y^n{e^{ - \frac{{{\beta _0}}}{{2k}}\left( {k_x^2 + k_y^2} \right)}}} {e^{i\tilde \theta }}.
	\label{1.38}
	\end{equation}
The phase $\tilde \theta $ is not required for the following considerations. It differs from Eq.~\ref{1.29} but is identical in Eqs.~\ref{1.38} and \ref{1.39}. Executing the integrals in Eq.~\ref{1.38} leads to a field description similar to Eq.~\ref{1.24}. (The result of Eq.~\ref{1.38} contains complex amplitudes. The conversion to real amplitudes and a detailed comparison of different mode representations can be found in~\cite{Pampaloni2004}.) The equivalent relation for the Laguerre-Gaussian beam reads as~\cite{Pampaloni2004}
	\begin{equation}
	\begin{gathered}
	\tilde E \propto \int\limits_{ - \infty }^\infty  {d{k_x}\int\limits_{ - \infty }^\infty  {dk_y}} \;\\
	{{\left( {{k_x} + i{k_y}} \right)}^n}{{\left( {{k_x} - i{k_y}} \right)}^{l + n}}{e^{ - \frac{{{\beta _0}}}{{2k}}\left( {k_x^2 + k_y^2} \right)}} {e^{i\tilde \theta }}.
	\end{gathered}
	\label{1.39}
	\end{equation}
Executing the integrals leads to a Laguerre-Gauss mode containing the Laguerre polynomial $L_n^l$, with the azimuthal mode number $l$, which describes the angular momentum of the beam, and the radial mode number $n$. The beam quality factor of a Laguerre-Gauss mode is given by 
	\begin{equation}
	{M_x} = {M_y} = {M_L} = 2n + \left| l \right| + 1.
	\label{1.40}
	\end{equation}
Eq.~\ref{1.39} contains imaginary wave components ($i{k_y}$), which means that these components are shifted relative to real part components by $\frac{\pi}{2}$ $(i{e^{ix}} = {e^{i(x-\frac{\pi}{2})}})$.
 This phase shift between orthogonal planes is characteristic for a spiral pattern of the wave front.\\ 
If we shift the phase of the imaginary amplitude component of Eq.~\ref{1.39} by $\frac{\pi}{2}$, so that the imaginary part becomes real, and transform into a coordinate system which is rotated around the $z$-axis by 45$^\circ$ in comparison to the original system, the Laguerre-Gauss mode Eq.~\ref{1.39} is transformed into an Hermite-Gauss mode Eq.~\ref{1.38}. In the rotated system, the relations
	\begin{equation}
	\begin{gathered}
  {{\tilde k}_x} = \frac{1}{{\sqrt 2 }}\left( {{k_x} + {k_y}} \right) \hfill \\
  {{\tilde k}_y} = \frac{1}{{\sqrt 2 }}\left( {{k_x} - {k_y}} \right) \hfill \\ 
\end{gathered}
\label{1.41}
\end{equation}
hold, thus the mode numbers $l$, $m$, and $n$ are related by $m = l + n$.\\
Knowing the relation of the mode numbers the beam quality factors before and after the mode converter can now be compared. Here we are interested in the projected 4D beam quality which yields for the Laguerre-Gauss mode
	\begin{equation}
	\begin{gathered}
	M_L^2 = {\left( {2n + \left| l \right| + 1} \right)^2}  \\ 
	    { = {\left( {2n + 1} \right)^2} + 2\left| l \right|(2n + 1) + {l^2}}  \\
			{ = M_{st}^2 + {l^2}}.
	\end{gathered}
\label{1.42}
	\end{equation}
As previously discussed the angular momentum is a correlated motion which contributes to the emittance, but it is not a fundamentally conserved quantity as a statistical emittance. In the last step of Eq.~\ref{1.42} the statistical beam quality factor $M_{st}^2 = {\left( {2n + 1} \right)^2} + 2\left| l \right|(2n + 1)$ is introduced. It represents the conserved quantity of motion. (${M_{st}}$ is the equivalent to the thermal emittance $\varepsilon_{th}$ in Kim's paper~\cite{Kim2003}.)\\
Using $m = l + n$ the mode numbers of the Hermite-Gauss mode can be written as
	\begin{equation}
	\begin{gathered}
  {M_x} = 2m + 1 = {M_L} + \left| l \right| \hfill \\
  {M_y} = 2n + 1 = {M_L} - \left| l \right| \hfill \\ 
\end{gathered} 
\label{1.43}
\end{equation}
and the 4D emittance follows as
	\begin{equation}
	{M_x}{M_y} = M_{st}^2
	\label{1.44}
	\end{equation}
The angular momentum of the beam is thus compensated and only the statistical emittance is preserved. Relations \ref{1.43} and \ref{1.44} correspond to the relations derived by Kim~\cite{Kim2003} for the beam emittance in case of the Derbenev transformation without making use of any mode description. The equivalence of these relations and of the Gouy phase with the Courant-Snyder phase found in the previous section, show that the pure mode converter and the Derbenev transformation are indeed identical.\\
The redistribution of the beam emittance is directly linked to the existence of an intrinsic angular momentum on one side, and the absence on the opposite side of the mode converter. The successful experimental demonstration (e.g.~\cite{Edwards2000,Edwards2001, Piot2006}) of the emittance redistribution and the tests of relevant properties of the converter optics proofs not only the existence, but it is also a qualitative measurement of the angular momentum. Independent measurements of the angular momentum, based on basic beam dynamics considerations (e.g.~\cite{Sun2004}), confirm these measurements. 

\section{Concluding remarks}
Derbenev's invention has led to numerous theoretical and experimental studies and a large number of proposals for its application. Besides control of the angular rotation, i.e., the kinetic angular momentum, in long solenoid sections, which is relevant for cooling applications~\cite{Burov2000}, the implementation of the beam adapter in electron storage rings, and the generation of beams with a large emittance ratio are being studied. Beams in electron storage rings develop typically a Hermite-Gaussian character, i.e., a large ratio of the emittance in the horizontal to the vertical direction due to differences in the damping rate related to synchrotron radiation. Round beam conditions can be generated in between two mode converters, which was proposed, e.g., to counteract beam-beam effects in an interaction region~\cite{Burov2002, Danilov1996} or for improved properties of radiation sources~\cite{Brinkmann2002}. The generation of charged beams -- electrons but also ions -- with large emittance ratio can be used, for example, to adapt beam properties to planar structures for the generation of radiation~\cite{Piot2012}, for advanced accelerators~\cite{Bell2018}, or for improving the injection efficiency into a synchrotron~\cite{Groning2017}. This list of examples is by no means complete. Moreover a general discussion on limitations and options for phase-space manipulations was stimulated~\cite{Kim2006}, also in conjunction with an emittance exchange concept in which one transverse and the longitudinal degree of freedom are involved~\cite{Emma2002}. Here $x-pz$ and $z-px$ correlations are introduced by means of a transverse deflecting RF structure in combination with dispersive sections. An interpretation of this beam adapter in terms of modes is however not yet possible.\\ 
The Courant-Snyder theory is a cornerstone of modern accelerator physics and it appears to be beneficial to apply it also to problems in laser physics. The optical functions lead to a consistent interpretation of the beam quality factors and the beam size of optical modes and it offers a simple approach for the description of light fields in transport lines. It has to be noted, that at this point only the very basic concept of the theory has been introduced. Many aspects, as for example the stability analysis of periodic systems, were not even touched. The general applicability of the Courant-Snyder theory to light beams was of course known, as it was to be expected that advantages emerge from its application. The equivalence of the Gouy and the Courant-Snyder phase however is somewhat surprising and establishes yet another connection between ray optics and wave optics.

\begin{acknowledgments}
I would like to thank Benno Zeitler and Dmitry Karlovets for careful proofreading, useful suggestions and stimulating discussions.  
\end{acknowledgments}

\appendix*
\section{Illustrative Examples}

Consider an arbitrary beam line of elements each of which is described by a linear transport matrix. The matrix transforms an incoming vector of particle coordinates $\left( {{x_i},\;{{x'}_i}} \right)$ into an outgoing vector $\left( {x,\;x'} \right)$.
	\begin{equation}
	\left( {\begin{array}{*{20}{c}}
  x \\ 
  {x'} 
\end{array}} \right) = \left( {\begin{array}{*{20}{c}}
  a&b \\ 
  c&d 
\end{array}} \right)\left( {\begin{array}{*{20}{c}}
  {{x_i}} \\ 
  {{{x'}_i}} 
\end{array}} \right).
\label{1.45}
\end{equation}
The coefficients of the matrix are in general $z$-dependent, so that the matrix represents a continuous function, i.e. it is a solution of Eq.~\ref{1.16}. Multiplication of the matrices of all elements allows describing the complete beam line.\\
Based on Eq.~\ref{1.45} $\left\langle {{x^2}} \right\rangle $, $\left\langle {{{x'}^2}} \right\rangle $ and $\left\langle {xx'} \right\rangle $ can be calculated which leads together with the relations to the optical functions to a corresponding matrix transforming the initial optical functions $\left( {{\beta _i},\;{\alpha _i},\;{\gamma _i}} \right)$ through the system 
	\begin{equation}
	\left( {\begin{array}{*{20}{c}}
  \beta  \\ 
  \alpha  \\ 
  \gamma  
\end{array}} \right) = \left( {\begin{array}{*{20}{c}}
  {{a^2}}&{ - 2ab}&{{b^2}} \\ 
  { - ac}&{ad + bc}&{ - bd} \\ 
  {{c^2}}&{ - 2cd}&{{d^2}} 
\end{array}} \right)\left( {\begin{array}{*{20}{c}}
  {{\beta _i}} \\ 
  {{\alpha _i}} \\ 
  {{\gamma _i}} 
\end{array}} \right).
\label{1.46}
\end{equation}
With this matrix also the phase advance through the system is determined.\\
The real part solution of Eq.~\ref{1.21} reads as
	\begin{equation}
	x = \sqrt {I\beta } \left( {\cos {\phi _i}\cos \Delta \phi  - \sin {\phi _i}\sin \Delta \phi } \right),
	\label{1.47}
	\end{equation}
with the initial phase angle ${\phi _i}$ and the phase advance through the system $\Delta \phi $. Note that Eq.~\ref{1.47} is valid for individual particle coordinates and for the average of a particle ensemble. The initial conditions are (cf. Figure~\ref{fig.2})
	\begin{equation}
	\begin{gathered}
  \cos {\phi _i} = \frac{{{x_i}}}{{\sqrt {I{\beta _i}} }} \hfill \\
  \sin {\phi _i} = \left( {{{x'}_i} + \frac{{{\alpha _i}}}{{{\beta _i}}}{x_i}} \right)\sqrt {\frac{{{\beta _i}}}{I}}  \hfill \\ 
\end{gathered} 
\label{1.48}
\end{equation}
 which leads to
	\begin{equation}
	x = {x_i}\sqrt {\frac{\beta }{{{\beta _i}}}} \left( {\cos \Delta \phi  - {\alpha _i}\sin \Delta \phi } \right) - {x'_i}\sqrt {\beta {\beta _i}} \sin \Delta \phi 
	\label{1.49}
	\end{equation}

\subsection{Propagation of an incoming angle or offset}
An incoming angle of the beam leads to trajectory excursions with local extrema at $\Delta \phi  = \left( n - \frac{1}{2}\right) \times 180^\circ $, $n = 1,\;2,\;3...$. The value of the local extreme is modulated with the local $\beta$-function. At these locations the incoming angle can be measured, while it can be corrected at locations where $\Delta \phi  = n \times 180^\circ$.
For an incoming offset the condition for an extreme is ${\alpha _i} = \tan \Delta \phi $, while zeros are found at  ${\alpha _i} =  - \cot \Delta \phi $. Thus extrema and zeros are also in this case separated by 90$^\circ$ phase advance.

 
\subsection{A simple imaging problem}
\begin{figure}[hhh]
\centering
    \includegraphics*[width=70mm]{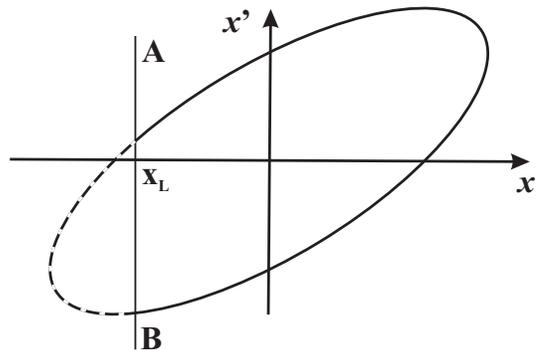}
    \caption{Phase-space with clipping aperture.}
\label{fig.3}
\end{figure}

Assume the beam clips at an aperture which is represented by a vertical line in phase-space, Figure~\ref{fig.3}. The condition for imaging the aperture is that all points forming the line $\overline {AB} $ are again at the same transverse position. Take two points on the line, e.g.  ${P_1} = \left( {{x_L},x'} \right)$ and ${P_2} = \left( {{x_L}, - x'} \right)$. The condition for aligning them is 
	\begin{equation}
	{x'_L}\sqrt {\beta {\beta _i}} \sin \Delta \phi  =  - {x'_L}\sqrt {\beta {\beta _i}} \sin \Delta \phi, 
	\label{1.50}
	\end{equation}
which requires that $\sin \Delta \phi  = 0$. The imaging condition is thus a phase-advance of $\Delta \phi  = n \times 180^\circ $. Note that the optical functions are derived for the undisturbed beam.

\subsection{Imaging of the reciprocal space}

 \begin{figure}[hhh]
\centering
    \includegraphics*[width=80mm]{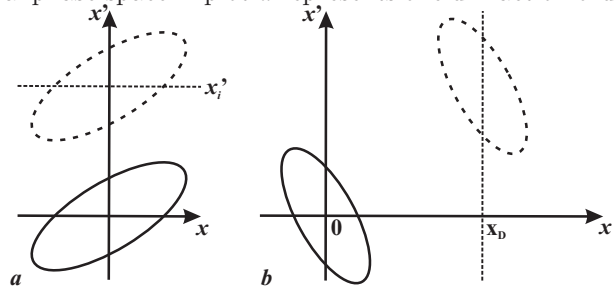}
    \caption{Diffraction imaging (schematic)}
\label{fig.4}
\end{figure}

A diffracting element generates additional structures in the momentum or angle coordinate of the phase-space.\\ 
Figure~\ref{fig.4} shows an example, where the replica of the initial phase-space in plot \textit{a} represents one diffraction order generated by a diffracting element at position \textit{a}. After traversing a section of a beam line the situation may look like displayed in \textit{b}. In the projection onto the $x$-axis two separated peaks are now visible. The condition for optimal imaging is that the distance of the diffraction peak to the central beam ${x_D}$ is large compared to the size of the central beam (or the diffraction peak). Thus the ratio 
	\begin{equation}
	\frac{{{x_D}}}{\sigma } = \frac{{{{x'}_i}\sqrt {\beta {\beta _i}} \sin \Delta \phi }}{{\sqrt {\beta \varepsilon } }} = {\sigma _i}{x'_i}\sin \Delta \phi 
	\label{1.51}
	\end{equation}
has to be maximized, which leads to the condition $\Delta \phi  = \left( n - \frac{1}{2} \right) \times 180^\circ $ for imaging of the reciprocal space. A large initial beam size (small uncorrelated divergence) increases the resolution power.

\end{document}